\documentclass[12pt]{article}
\usepackage[]{graphicx}
\usepackage{natbib}
\begin{document}

\begin{center}

{\large FU Orionis: A Binary Star?}\\
\vspace{0.5cm}
Hongchi Wang\\
Max-Planck-Institut f\"{u}r Astronomie, K\"{o}nigstuhl 17, D-69117 Heidelberg,
Germany; email: wang@mpia-hd.mpg.de\\
and\\
Purple Mountain Observatory, Academia Sinica, Nanjing 210008, PR China\\
\vspace{0.5cm}
D\'aniel Apai, Thomas Henning, and Ilaria Pascucci\\
Max-Planck-Institut f\"{u}r Astronomie, K\"{o}nigstuhl 17, D-69117 Heidelberg,
Germany\\
\vspace{1.0cm}
{\large Abstract}\\
\end{center}
\hspace{18pt}By using the ALFA adaptive optics system at the 3.6m telescope of the Calar 
Alto Observatory 
we detected a faint red star in the apparent vicinity of FU
Ori, the prototype of the FUor outburst stars. Independent confirmation
of the detection is obtained from archival PUEO/CFHT images. 
The separation between the companion candidate and FU Ori 
is 0.$^{\prime\prime}$50 and their brightness contrast is around 4 
magnitudes. We discuss the possible nature of the newly detected star based on 
near-infrared photometry and its proper motion 
relative to FU Ori. The photometric data are consistent with a nearby late-type main 
sequence star, a background giant star, 
and a pre-main sequence star. 
On the basis of the proper motion and the stellar surface density in the 
direction towards FU Ori, we argue that the probabilities of the first 
two options are very low.

\vspace{0.5cm}
{\it Subject headings:} binaries: general --- stars: individual (FU Ori) --- stars:
  pre-main-sequence --- techniques: high angular resolution

\section{Introduction}

\hspace{18pt}Recent studies of low-mass stars in the star-forming regions 
Taurus \citep{Gullbring98} and 
Orion \citep{Robberto03} indicate surprisingly 
low accretion rates between $\dot{M}\simeq 10^{-7} - 10^{-10} M_{\odot} yr^{-1}$.
These accretion rates are far too low to build up solar-type stars on
timescales of the order of 10$^{6}$ yr,
as it is expected in the current picture of low-mass star formation 
(see \citealt{Lada99} and references therein). An elegant alternative to the 
quiescent, steady accretion can be the occurence of periodic, short events 
of intense accretion ($\dot{M}\simeq 10^{-4} M_{\odot} yr^{-1}$), 
during which a significant fraction of the disk mass can be accreted \citep{Hartmann96}.
The well-known FUor objects \citep{Herbig66,Herbig77} may represent this intense 
accretion phase of otherwise quiescent T Tauri stars \citep{Hartmann96}. 
Thus, if FUor outbursts are typical to {\it all} T Tauri stars, these outbursts
might be the dominant way of mass accretion. 

Models based on the enhanced accretion 
rates are in general successful in explaining the observed properties of the 
FUor outbursts \citep{Hartmann85,Kenyon88}, but the causes of the 
onset of the rapid accretion are still disputed.
Suggested possibilities include thermal instabilities in the disk 
(see, e.g., \citealt{Lin85}; \citealt{Kawazoe93};
\citealt{Bell94}) and perturbations from close companions 
\citep{Bonnell92,Clarke96}, but up to now none of these
 explanations could be confirmed observationally.
 
 However, earlier high-angular resolution 
 observations led to the detection of companions 
 for the FUor objects L1551 IRS 5 \citep{Rodriguez98} and Z CMa
 \citep{Koresko91}, with a separation of 45 and 93 AU respectively. 
 Moreover, \citet{Kenyon93} 
 found RNO 1B and 1C to be a FUor binary.
Infrared long-baseline interferometry 
shows that FU Ori might have a close companion with a separation 
as small as 0.35 $\pm$ 0.05 AU, although these observations could be explained 
more naturally with the presence of a circumstellar disk \citep{Malbet98}. 

In this letter we present adaptive optics imaging of the young pre-main sequence
star FU Ori and the detection of a faint red star in its apparent vicinity.
Further results from our imaging campaign to search for companions of young stars 
were reported by \citet{Apai03}.
 
In order to avoid confusion, throughout the letter we use the name FU Ori for 
the prototype star ($\alpha = 05^h45^m22.^s6, \delta =
+09^\circ04^{\prime}12^{\prime\prime}$), while the term FUor stands for the 
class of stars named after FU Ori.

\section{Observations and data reduction}

\hspace{18pt}On 2002 October 27 we observed FU Ori and the star HD
38224 as PSF-reference using the near-infrared camera Omega Cass 
mounted at the 3.6m telescope at the Calar Alto Observatory, Spain.
The plate scale used in our observations 
was 0.$^{\prime\prime}$038 pixel$^{-1}$. The weather conditions changed
between good and excellent, with mean optical seeing of 0.$^{\prime\prime}$7. 
The observations aimed to enhance the contrast of the imaging system by the 
subtraction of a reference PSF from the target
object. In order to compensate for the temporal variations of the PSF due to
the changes in the atmosphere and in the optical system, we applied 
short observing {\it cycles}, alternating between the target and 
the PSF reference star.
 
Each {\it cycle} consisted of four dithering positions around the target
with roughly 83 seconds (98 $\times$ 0.842 second) spent at each position.
Thus, the total on-source integration time of a cycle was 332 seconds, both in
J and Ks filters. In the J filter we carried out two cycles on FU Ori, two 
cycles on the PSF star and two additional cycles on FU Ori. These cycles were 
observed immediately one after the other to minimize the PSF variations. 
Due to the approach of sunrise, in Ks band we did only one cycle on FU Ori, 
one on the PSF star and one again on FU Ori. The resulting total integration 
times were 1320~s in J and 660~s in Ks on FU Ori, and 660~s in J and 330~s in Ks 
on the PSF star.

The basic data reduction was conducted in the standard fashion with 
flat field correction and bad pixel removal. The sky frame for each cycle was 
obtained by taking the minimum of the images at different dithering 
positions. This sky frame was subtracted from each individual image of a 
given {\it cycle}. Following this, the frames from a single {\it cycle} were
combined into a mosaic image. The relative shifts of the individual 
frames were determined by cross-correlating the images.
  
In order to enhance the contrast between a possible companion and FU Ori, we
subtracted a brightness-scaled and positionally aligned PSF, known from the
PSF reference star (for details of the method see \citealt{Pantin00}).  

\section{Results}

\hspace{18pt}Our data reduction procedure yielded the PSF-subtracted J and Ks band images
of the star FU Ori (see Fig.~\ref{fig1}). The central part ({\bf
$r<0.^{\prime\prime}4$}) 
is heavily contaminated by the PSF-subtraction residuals and speckle 
phenomena. The residuals originate from the imperfect PSF subtraction 
due to the temporal variations of the PSF. The speckle pattern differences lead 
to the {\it speckle boiling} \citep{Racine99}, resulting in a dotted noisy 
pattern. South-east from the FU Ori residuals an over-subtracted
(negative) star is visible. This faint star is a previously unknown 
(visual) companion of the PSF-reference star HD 38224. This (visual)
companion star is also identifiable on the reduced 
(non-PSF-subtracted) images of HD 38224, but it becomes very evident
after the subtraction. We note that in the Ks band the first Airy-ring
of this star is well visible.

The major result of our observations is the detection of a
previously unknown star in the FU Ori images to the south of FU Ori 
(see Fig.~\ref{fig1}). In the following we will refer to this star 
as FU Ori S. The following facts exclude the possibility of FU Ori S being 
an artifact:

1. The star has been detected in both the J and Ks bands at the same location 
(see panels A and B in Fig.~\ref{fig1}):
The position difference is less than 0.$^{\prime\prime}$01 and the position angle
difference is only 2.$^\circ$8 -- within the errors of the position 
determination (0.$^{\prime\prime}$03; 3$^\circ$).

2. The star shows an Airy-ring pattern in the Ks band image,
like the (visual) companion of the PSF reference star.

3. The star is also detected at the same position, when another
PSF star (HD 201731 in the J-band, XY Cep in the Ks band)
from the same night is used as PSF reference. Although these
PSF-stars were observed several hours before FU Ori, subtracting
them from FU Ori reveals again the existence of FU Ori S.
 
4. Archival data from the AO system PUEO mounted on the
Canada-France-Hawaii Telescope (shown in Fig.~\ref{fig2}) 
provides independent confirmation for the existence of FU Ori S.
The co-added image was obtained through the K continuum filter 
($\lambda_c = 2.260 {\mu}m, \Delta\lambda
= 0.060 {\mu}m$) with a total exposure time of 11.2 seconds.

With respect to FU Ori, we derive from our ALFA images a position angle of 
$160.^\circ8 \pm 3^\circ$ and a separation of $0.^{\prime\prime}50 \pm
0.^{\prime\prime}03$ 
(linear separation = 225 $\pm$ 14 AU) for FU Ori S. 
The positional error given here is dominated by the centering 
uncertainties of FU Ori S (around 0.5 pixels). 

Accurate photometry of FU Ori S suffers from
the PSF-subtraction residuals and the speckle noise. In order to 
reduce these effects, we used a photometric aperture as small as
0.$^{\prime\prime}19$ ($\simeq$ 1.6 $\times$ FWHM).
Our photometric conversion factors were deduced by applying this
aperture to the stars FU Ori, HD 38224, and HD 201731 and
using their 2MASS Point Source Catalogue fluxes. The error on
the conversion factor in the J band is estimated to be 0.22 mag. 
In Ks band, due to 
the saturation of FU Ori and HD 201731 we can only
derive the conversion factor for HD 38224. 
As the AO system performs better in the Ks band than in the J band, the 
conversion factor error in the Ks band should be smaller than that in the J 
band. To be conservative we adopt the error of the conversion factor 
in the Ks band to be the same as that in the J band. 
Another major uncertainty of the aperture photometry 
of FU Ori S comes
from the speckle pattern. To estimate its influence on
our photometry, we have integrated the flux of speckle noise in 18 positions
over an aperture with the same diameter as used for FU Ori S. 
Comparing these fluxes to that of FU Ori S, we estimate an error
due to speckle noise of 0.11 mag and 0.08 mag in J and Ks bands, respectively.

Based on the aperture photometry and the described error estimates,
we derive an apparent brightness of 
J= 10.65 $\pm$ 0.25 mag and Ks= 9.64 $\pm$ 0.23 mag for FU Ori S. 
The apparent brightness of FU Ori is J = 6.519 $\pm$ 0.023 mag, 
K = 5.159 $\pm$ 0.020 mag \citep{Cutri03}. Therefore, the brightness 
contrast between FU Ori and FU Ori S amounts to 4.13 and 4.48 mag in the 
J and Ks bands, respectively.

\section{Discussions}

\hspace{18pt}The result of our AO observations is the identification of
a faint star close to FU Ori, the prototype of the FUor stars. In the following 
we assume 450 pc as the distance of FU Ori. Our photometry shows the unusually 
red color of FU Ori S (J - Ks = 1.01 $\pm$ 0.34) which is similar to that of 
FU Ori itself (J - K = 1.36 $\pm$ 0.03).
Although unambiguously determining the nature of FU Ori S from only two 
near-infrared fluxes is not possible, valuable conclusions can be drawn from 
its photometric properties.
In this section, we first discuss the probability of FU Ori S being an unrelated 
field star, then the possibility of FU Ori S being a late-type main sequence (MS) 
star, a giant/supergiant or a pre-MS star with Ks-band infrared excess. 

The very vicinity of FU Ori S to FU Ori suggests that FU Ori S is
most probably a companion of FU Ori. The 2MASS Point Source Catalogue 
\citep{Cutri03} shows that the FU
Ori region is not densely populated: in a region with a radius of
8$^\prime$  centered on FU Ori there are only 6 stars brighter than FU Ori S 
(Ks = 9.64 mag) in the K band. In analogy to earlier binarity surveys, 
we would consider any bright star closer than $\sim$ $2.^{\prime\prime}5$ to 
FU Ori as a companion. Therefore, the probability for a bright
field star to be considered as a companion by chance is
less than ${6 \over (8 \times 60)^2 \pi } \times 2.5^2 \pi =
1.6 \times 10^{-4}$. This estimate strongly supports the view
that FU Ori S is a companion of FU Ori.

The photometry and the proper motion data give further constraints on
 the nature of FU Ori S. If it is a MS star, its spectral type -- assuming
 no extinction -- will be later than K4 to match its red color. A simple 
 comparison of the absolute brightness of MS stars to the apparent 
 brightness of FU Ori S sets the upper distance limits of 120 pc for K4 and 50 
 pc for M6 spectral types. Such a
 nearby star, however, is likely to display a measurable proper motion over
 the 5 years time base between the ALFA and PUEO observations. For
 example, from the Hipparcos catalogue \citep{Perryman97} the average annual 
 proper motion of stars located at distances
 between 50-120 pc and within a radius of 10$^\circ$  around FU Ori is 38 
 mas yr$^{-1}$. In contrast, the absolute proper motion of FU Ori itself is
 less than 6 mas yr$^{-1}$ (see Tycho catalogue \citep{Hog97}). Thus, if FU Ori
 and FU Ori S are unrelated, their relative
 separation should have changed by an amount of $0.^{\prime\prime}19$ over 5
 years, a clearly detectable shift in the AO images. 
Comparing the PUEO position to that derived from ALFA we found a position
 difference of 10 $\pm$ 48 mas (the centering error of the PUEO images
 is estimated to be 1 pixel). This gives a proper motion limit of 
 2 $\pm$ 10 mas yr$^{-1}$ and indicates that FU Ori S is unlikely to be a nearby 
 MS star.

 Such a very small relative proper motion can be consistent with a
 star co-moving with FU Ori or a distant background object.
 Considering the option that FU Ori S is a background giant -- assuming
 the case of no extinction for simplicity -- we obtain that giants with
 spectral types between K1 and M7 at distances between 2 to 34 kpc are
 consistent with the photometry. Additional reddening would
 decrease the distances but would not change the overall picture. 
 
 The third and most exciting possibility is that FU Ori S is a pre-MS star 
 associated to FU Ori itself. A simple comparison to the isochrones of 
 \citet{Baraffe98}
 shows that the observed fluxes are in an overall good agreement with the 
 predicted magnitudes of a $\sim$ 1.1 M$_\odot$ star of the age of $10^{6}$ yr at 
 the 450 pc distance of 
 FU Ori (J$_{\rm BCAH}$=10.61 mag and K$_{\rm BCAH}$ = 9.80, for
 [M/H] = 0, Y = 0.275, and L$_{mix}$ = H$_{p}$). Apparently FU Ori S is somewhat redder than
 these pre-MS isochrones. This could be due to a reddening of A$_v$ = 1.1 mag 
 and/or the infrared excess arising from warm circumstellar material. 
 We point out, that assuming FU Ori S as a pre-MS star provides a natural 
 explanation for the observed brightness, color and the lack of proper motion 
 relative to FU Ori. Assuming a physical association between FU Ori S and FU Ori, 
taking the mass of FU Ori S to be 1.1 M$_\odot$ and the mass of FU Ori to be 
0.5 M$_\odot$, the typical mass of T Tauri stars, the primary in the 
FU Ori binary system is in fact FU Ori S, rather than FU Ori itself. 
Moreover, if we assume circular orbits in the plane of sky, the above masses 
and the linear separation (225 $\pm$ 14 AU) yield an orbital period of $\sim$ 2700 yr.
 
 In the case of FU Ori S being physically associated with FU Ori, the properties
of the FU Ori binary matches well with those used in the model of 
\citet{Bonnell92}. For example, in their 1G and 1O simulations the mass
ratios are 0.67 and 0.42, respectively, and the intervals between the companion 
periastron passes are around 2500 and 2100 yr \citep[see][Figs 2 and
3]{Bonnell92}. In contrast, the mass ratio and the companion 
separation of the FU Ori binary candidate differs strongly from those used 
in the \citet{Clarke96} model in which a companion of mass of 10$^{-2}$ M$_\odot$
at a very close distance ($\sim$ 0.1 AU) to the primary is exploited
to trigger the enhanced accretion. We note that the detection of more and more 
companions of FUor stars points to the possibility that they may play an 
active role in triggering the outbursts.
 
 \section{Conclusions}

 \hspace{18pt}Our AO-assisted near-infrared imaging led to the detection of 
 a red point source south of the well-known outburst star FU Ori.
 We discuss the possible nature of this object and conclude that
the observed properties are consistent with either a pre-MS star 
at the distance of FU Ori or -- with much smaller probability -- 
a red background giant star. We argue that mid-resolution optical 
spectroscopy can discriminate between these cases.

 As the FUor outbursts may be of central importance for star formation,
 a possible companion to the prototype object FU Ori can have a
 major impact on the outburst models. In particular, the close 
 fly-by of the companion could lead to disk perturbations, therefore, 
 providing an external triggering of the outbursts of FUor stars.
 
We would like to thank the staff members of the Calar Alto Observatory for
their support during the observations. We are grateful to Ch. Leinert 
for his useful comments. H. Wang acknowledges the support by NSFC
grants 10243004 and 10073021.

\clearpage

\begin{figure}
\caption[f1a.eps and f1b.eps]{PSF-subtracted images of FU Ori, (a) in J and (b) in Ks band. 
North is up and the east is to the left. The positions of FU Ori, FU Ori S, and 
the visual companion of the PSF reference star are indicated with numbers 1-3 in
(b). The image scale is marked in (b).}
\label{fig1}
\end{figure}

\begin{figure}
\caption{CFHT PUEO image of FU Ori through the K continuum filter ($\lambda_c = 2.260 {\mu}m, \Delta\lambda
= 0.060 {\mu}m$). The observations were made on 1997 December 22. 
The total exposure time is 11.2~s and no PSF-subtraction was applied. 
The position of FU Ori S is indicated.}
\label{fig2}
\end{figure}

\end{document}